\documentclass{article}
\usepackage{graphicx} % Required for inserting images
\usepackage{physics}
\usepackage{amsmath,amssymb}
\usepackage{slashed}
\usepackage{bbold}
\usepackage[style=nature, backend=bibtex]{biblatex} %Imports biblatex package
\usepackage{geometry}
\usepackage{authblk}
\usepackage{abstract}
\usepackage{hyperref}
\usepackage{xurl}
\hypersetup{breaklinks=true}

\def\correspondingauthor{\footnote{cameron.cianci@uconn.edu}}

\title{Security of One-Way Entanglement Purification with Quantum Sampling Against a Restricted Adversary}
\author[1]{Cameron Cianci\correspondingauthor{}}
\affil[1]{Physics Department, University of Connecticut, Storrs, CT 06269, USA}
\date{}

\begin{document}

\maketitle

\begin{abstract}
Entanglement purification protocols promise to play a critical role in the future of quantum networks by distributing entanglement across noisy channels.  However, only the security of two-way purification protocols have been closely studied.  To address this, we propose a one-way entanglement purification protocol which utilizes quantum sampling and prove its security against an adversary restricted to single qubit Pauli gates.  This is done through leveraging the equivalence of one-way entanglement purification protocols with error-correcting codes.  To prove the security of this protocol, we first use the quantum sampling framework introduced by Bouman and Fehr to estimate the Hamming weight of the qubits which passed through the channel and then use the estimated relative Hamming weight $\omega$ to determine the amount of interference that Eve has subjected to the quantum channel.  Since Eve is restricted to single qubit Pauli gates, the number of applied gates can be directly estimated using the Hamming weight.  Estimating the number of adversarial single qubit gates, allows us to perform error correction and disentangle the logical qubit from Eve with probability $1-\epsilon_{qu}^\delta$.  Since this protocol allows communication only in one direction, the distance of the code must be decided before transmission, and therefore Bob will be forced to abort the protocol if he finds that Eve has applied more gates than the code can correct. One-way protocols may find use when communication is limited, or when we desire to decrease latency compared to the multiple rounds of communication needed in two-way protocols.  Further research may investigate the security of this protocol against arbitrary single or multi-qubit gates to obtain security guarantees against a more general adversary.
\end{abstract}

\section{Introduction}
Entanglement is a novel computational resource in quantum information science with no similar classical resource.  This resource has found numerous uses throughout quantum information science.  Two parties with entangled qubits can transmit perfectly secure quantum information through quantum teleportation \cite{4804520, Bennett_19} or perfectly secure classical information through superdense coding \cite{PhysRevA.71.044305}.  Parties with access to entanglement can use quantum correlations to succeed at games such as "magic squares" more than is classically possible \cite{Leditzky_2020}.  Quantum entanglement has also found use in quantum processes such as distributed quantum computation \cite{https://doi.org/10.48550/arxiv.2212.10609} and quantum cookies \cite{Bennett_1996}.  Motivated by these uses, the topic of how to securely distribute quantum states has recently gained interest, including through investigation of quantum secure direct communication \cite{PhysRevLett.89.187902,  https://doi.org/10.1002/que2.26, 9129730} and secure quantum dialogues \cite{Ye_2014, Ye_2013}.

In contrast to the quantum secure direct communication and quantum dialogue protocols above which attempt to distribute entanglement through bell pairs, the protocol proposed in this paper will attempt to distribute entanglement through error correcting codes.  Due to this focus on distributing entanglement, we will allow Alice and Bob the classical resource of a shared secret key prior to the start of the protocol.  In this protocol, we will also utilize entanglement purification protocols, which were historically brought about to distill high fidelity entangled states quantum states travelling through a noisy channel \cite{Bennett_199}.  However, since an eavesdropper can be viewed as a source of noise in the quantum channel, entanglement purification protocols also work to remove the effect of an adversary on the transmitted message. Entanglement purification protocols have been investigated for use in high fidelity quantum communication \cite{Pan_2001}, and previously explored two-way protocols promise strong security for transferring quantum information \cite{PhysRevA.66.032302}.  

\subsection{Entanglement Purification Protocols}
There are two different classes of entanglement purification protocols, one-way and two-way protocols \cite{Bennett_1996}. Two-way entanglement purification protocols allow for communication between Alice and Bob after qubits pass through the quantum channel.  Since Alice and Bob can conditionally apply gates to their systems based off each other's measurement results, two-way entanglement purification protocols are in general stronger than one-way protocols.  For example, two-way purification protocols can allow for Alice and Bob to purify their qubits from a $50\%$ depolarizing channel, which one-way protocols cannot correct \cite{Bennett_1996}.  Two-way purification protocols have been previously proven secure through demonstrating that a two-way protocol can disentangle the purified system from an external system or eavesdropper \cite{PhysRevA.66.032302}.  This posits the question if one-way protocols can similarly be proven secure. 

In comparison to two-way protocols, one-way entanglement purification protocols do not allow for communication between parties after the message is sent.  Restriction to one-way communication incidentally makes these protocols equivalent to quantum error correction since Alice and Bob can be time-like separated \cite{Bennett_1996}.  Due to this, we can use quantum error-correcting codes to evaluate the security of one-way entanglement purification protocols \cite{D_r_2007}.  The correspondence between one-way entanglement purification protocols and error-correcting codes has a prior basis in the literature, as it has previously been used by Shor and Preskill to prove the BB84 QKD protocol secure \cite{Shor_2000}. Investigating the security of one-way protocols may be useful in scenarios where communication between two parties is limited, or when we desire to decrease the latency of quantum communication compared to the multiple rounds of classical communication needed between Alice and Bob in two-way protocols.

In one-way entanglement purification, Bob is forbidden from sending messages to Alice.  This presents a problem for quantum sampling, as this restriction requires Bob to know the states in which the sampling qubits were prepared in order to perform sampling and determine the Hamming weight of the sampling qubits.  If Alice naively announces the prepared sampling states over a classical channel, then Eve can simply intercept all the qubits and use this information to resend identical qubits to Bob, circumventing the quantum sampling procedure.  Fundamentally, this problem occurs because in one-way protocols the actors Bob and Eve are symmetric \cite{Bennett_1996}.  To solve this problem, we will assume that Alice and Bob share a secret classical key, breaking this symmetry.  

A shared classical key will allow Alice to send a secure classical message to Bob.  This message will contain many critical pieces of information for the protocol, including the prepared sampling states, the permutation which Alice applied to the transmitted qubits, and the distance of the error-correcting code which Alice employed.  At this point in the protocol, Bob will be able to perform quantum sampling and estimate the relative Hamming weight $\omega$ of Eve's attack.  Bob will then be able to use the relative Hamming weight to estimate the number of gates Eve has applied to the logical qubit, if we assume Eve is restricted to single qubit Pauli gates. Finally, Bob can calculate if the error-correcting code distance is large enough to disentangle Eve.  If the code is sufficiently large, then Bob performs error correction and keeps the resulting logical qubit. Otherwise, the code distance is too small and Bob aborts.  

\section{Sampling}
Before presenting the protocol we should first familiarize ourselves with quantum sampling.  The quantum sampling framework used here was introduced by Bouman and Fehr \cite{https://doi.org/10.48550/arxiv.0907.4246}, and we refer to their paper for a more rigorous introduction.  This section is intended to be a broad overview to the framework put forth there.  

Generically, sampling allows for an individual to learn information about a population through measuring a subset of that population. Classical sampling has been well studied \cite{Thompson_2012a}.  However, due to entanglement in quantum systems it is not obvious how classical sampling strategies could be extended into quantum systems.  Bouman and Fehr's quantum sampling framework addresses this by allowing for classical sampling strategies to be used in quantum systems.  The sampling framework outputs an estimate of the Hamming weight of a quantum system.  The definition of the Hamming weight is extended in this framework to entangled states and states in superposition \cite{https://doi.org/10.48550/arxiv.0907.4246}.  

From Bouman and Fehr's quantum sampling framework we will be interested in two quantities: the estimated relative Hamming weight $\omega$ of the message qubits and the error bound of the quantum sampling process $\epsilon_{qu}^\delta$.  In Sections 5 and 6 we will use the relative Hamming weight $\omega$ to estimate the number of gates applied by an eavesdropper, and we will use the error bound $\epsilon_{qu}^\delta$ to estimate the probability that the sampling strategy has failed and Eve has gained access to the transmitted logical qubit without detection.

\subsection{Classical Sampling}
As an introduction to sampling, let us consider a classical string $q = (q_1, q_2, ... q_n) \in \{0,1\}^n$.  The Hamming weight of this string is defined as $wt(q) = |\{ i | q_i \neq 0 \}|$, or in simpler terms, the Hamming weight is the number of nonzero characters in this string.  The role of sampling is to use a substring $q_t$ to estimate the Hamming weight of the remaining string $q_{\bar{t}}$. The sampling strategy employed will select a completely random subset of $q$ to determine $q_t$. With the sampled substring $q_t$, the relative Hamming weight $\omega(q_t) = \frac{wt(q_t)}{N}$ will be used as an estimate of the relative Hamming weight of the remaining string $\omega(q_{\bar{t}}) = \frac{wt(q_{\overline{t}})}{M}$, where $N$ is the number of sampling qubits and $M$ is the number of message qubits.

To find the failure probability of the sampling procedure, we will start by considering the set of all substrings that would output a relative Hamming weight which is $\delta$-close to the true relative Hamming weight.

\begin{equation}
    B_t^\delta = \{ \textbf{q} \in \mathcal{A}^n : \abs{\omega(q_{\overline{t}}) - \omega(q_t)} < \delta \}
\end{equation}

This is the set of all substrings for which the sampling procedure will succeed.  Through considering this set, it is clear that the probability that the sampling procedure fails, producing an estimate greater than $\delta$ from the true Hamming weight, which is equal to the probability that the string is not in the set $B_t^\delta$.

\begin{equation}
    \epsilon_{cl}^\delta = \max_{q \in \mathcal{A}^n} Pr[q \notin B_t^\delta]
\end{equation}

Assuming we are randomly sampling $k$ entries \cite{https://doi.org/10.48550/arxiv.0907.4246}, we find,

\begin{equation}
    \epsilon_{cl}^\delta < 4\exp{-\frac{1}{3}\delta^2 k}
\end{equation}

Therefore, classical sampling is able to estimate the relative Hamming weight $\omega(q_t)$ of a string which is $\delta$-close to the true relative Hamming weight $\omega(q_{\bar{t}})$, with probability $1-\epsilon_{cl}^\delta$.  From this point on we will simply refer to the estimated relative Hamming weight as $\omega$.

\subsection{Quantum Sampling}
The quantum version of sampling naturally extends from classical sampling in Bouman and Fehr's framework, except that sampling is performed in both the $X$ and $Z$ bases.  Classical sampling along these two non-orthogonal bases allows for us to estimate the Hamming weight of the quantum system while still using well-studied classical sampling methods.  The main result of Bouman and Fehr's paper shows that the error bound of a quantum sampling protocol is simply the square root of the error bound of the underlying classical sampling protocol utilized \cite{https://doi.org/10.48550/arxiv.0907.4246}.

\begin{equation}
    \epsilon_{qu}^\delta = \sqrt{\epsilon_{cl}^\delta}
\end{equation}

Using this, we find the quantum error bound for a randomly sampled substring to be,

\begin{equation}
    \epsilon_{qu}^\delta < 2\exp{-\frac{1}{6}\delta^2 k}
\end{equation}

Through this equation, Bouman and Fehr's framework allows for classical sampling methods to be applied to quantum systems.  This finding has allowed this framework to aid in proving the security of Quantum Key Distribution \cite{https://doi.org/10.48550/arxiv.0907.4246} and Quantum Random Number Generators \cite{yao2020quantum, Krawec2019}, as well as in deriving lower bounds on the quantum conditional min entropy of high dimensional systems \cite{9174330}.  

\begin{center}
    \includegraphics[scale = .35]{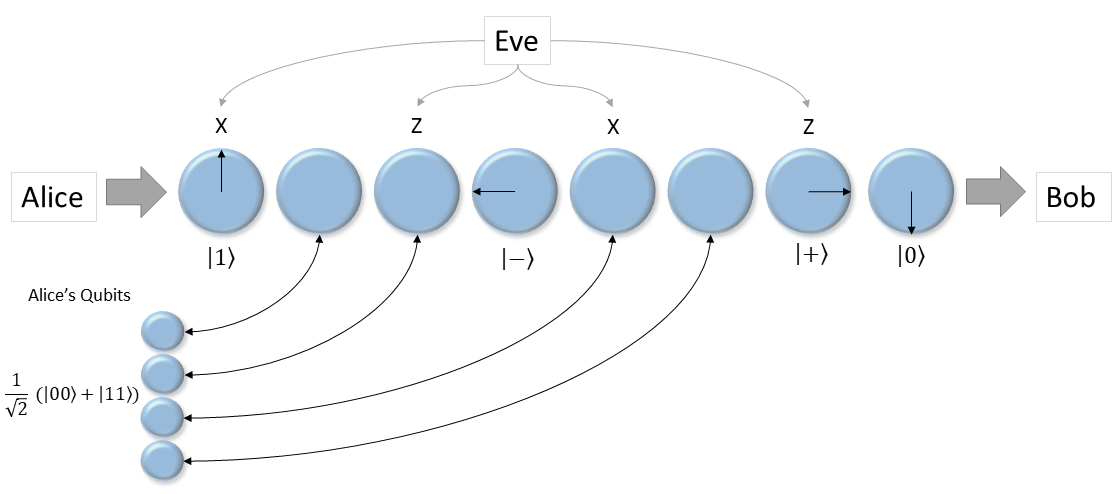}
    
    \textbf{\footnotesize Figure 1. Sampling and message qubits sent over a quantum channel.  Eve applies Pauli gates randomly to the message and sampling qubits.  Bob can estimate the number of Pauli gates applied to the message qubits through the sampling procedure.}
\end{center}

To illustrate quantum sampling with an example, consider the situation in Figure 1. 
Focusing on the first qubit, Alice prepared a sampling qubit in the $\ket{1}$ state and sent it to Bob over a quantum channel.  However, Eve tampered with the qubit in the channel, and when Bob measured this qubit he found it in the $\ket{0}$ state.  In this case, Bob can deduce that an error has occurred.  Now considering all the sampling qubits sent, Bob can obtain a more accurate estimate of the influence of the quantum channel.  With this information, when Alice sends additional message qubits along with these sampling qubits, Bob can use the quantum sampling to estimate the total error of this quantum message in the form of the Hamming weight, $M\omega$.

When analyzing the proposed protocol in Section 5, $\omega$ will be used to determine the number of gates Eve has applied, and $\epsilon_{qu}^\delta$ will be used to determine the failure probability of the protocol.  The value of $\delta$ will be determined in Sections 5 and 6 by the Hamming weight, code distance, and gate set given to Eve.

\section{Estimating Eve's Interference}

We will now use the Hamming weight, along with restrictions on Eve's available gate set, to estimate the number of gates Eve has applied.  Let us first consider the example of Eve tampering with the quantum sampling procedure in Figure 1.  Alice began by preparing sampling qubits in the states $\ket{0}$, $\ket{1}$, $\ket{+}$, and $\ket{-}$ and sent these sampling qubits along with some message qubits to Bob.  After Bob received these qubits, he measured the sampling qubits in the same basis Alice prepared and estimated the relative Hamming weight $\omega$ of the remaining qubits. Bob can use this relative Hamming weight $\omega$ to gain insight into the possible attacks Eve could have performed as follows.  

The Hamming distance is defined as the number of qubits which would be measured by Bob in an orthogonal state as compared to the state which Alice prepared.  This change of state is caused by Eve's interference in the quantum channel, shown below by the operator $E$.  This implies a definition of the relative Hamming weight as follows,

\begin{equation}
    \omega = \frac{1}{4}\abs{\bra{1}E\ket{0}}^2 + \frac{1}{4}\abs{\bra{0}E\ket{1}}^2 + \frac{1}{4}\abs{\bra{+}E\ket{-}}^2 + \frac{1}{4}\abs{\bra{-}E\ket{+}}^2
\end{equation}

Extending this equation to multiple qubits gives the Hamming weight for the sampling qubits as,

\begin{equation}
   N\omega = \sum_i^N \frac{1}{4}\abs{\bra{1}E_i\ket{0}}^2 + \frac{1}{4}\abs{\bra{0}E_i\ket{1}}^2 + \frac{1}{4}\abs{\bra{-}E_i\ket{+}}^2 + \frac{1}{4}\abs{\bra{+}E_i\ket{-}}^2
\end{equation}

By multiplying the relative Hamming weight by the number of message qubits $M$, we can determine the estimated Hamming weight of the message qubits,

\begin{equation}
   M\omega = \frac{M}{N} \sum_i^N \frac{1}{4}\abs{\bra{1}E_i\ket{0}}^2 + \frac{1}{4}\abs{\bra{0}E_i\ket{1}}^2 + \frac{1}{4}\abs{\bra{-}E_i\ket{+}}^2 + \frac{1}{4}\abs{\bra{+}E_i\ket{-}}^2
\end{equation}

Given $\omega$, we will use this equation to gain insight into the operators $E_i$ in Sections 5 and 6.  In these sections we will find that knowledge of the gate set and the Hamming weight of the message qubits $M\omega$ can be used to estimate the total number of gates applied.

\section{The Protocol}
With quantum sampling and the Hamming weight of quantum systems better understood, we are now in a position to state the proposed one-way entanglement purification protocol, which is as follows,

\begin{enumerate}
    \item Alice chooses a code distance $d$ and encodes a logical qubit with this distance.  She entangles this logical qubit with a local qubit, which she will keep.
    \item Alice then concatenates many sampling qubits to this logical qubit and permutes her quantum registers.  She sends all these qubits through the quantum channel.
    \item Alice sends a classical message to Bob using a previously distributed shared secret key, informing him of the permutation, sampling states, and code distance.
    \item Bob receives both the classical message and qubits from their respective channels.
    \item Bob uses the classical information given by Alice to locate the sampling qubits and perform quantum sampling, obtaining an estimate of the relative Hamming weight $\omega$.  
    \item Using the estimated Hamming weight $M\omega$, Bob can additionally estimate the number of operations applied to the logical qubit.  If the number of operations is less than that allowed by the code distance, then Bob performs error correction and keeps the logical qubit.  Otherwise, Bob aborts.
\end{enumerate}

Bob's prediction is correct with probability $1-\epsilon_{qu}^\delta$.  Bob can determine the value of $\delta$ as is shown in the next section.

\section{Security Against Pauli Gates}
We will now prove the security of this protocol when Eve's gate set is restricted to single qubit Pauli gates.  Starting from equation 8, with Eve's attacks restricted to single qubit Pauli gates $E_i \in \{X_i, Y_i, Z_i\}$,

\begin{equation}
   M\omega = \frac{M}{N}\sum_i^N \frac{1}{4}\abs{\bra{1}E_i\ket{0}}^2 + \frac{1}{4}\abs{\bra{0}E_i\ket{1}}^2 + \frac{1}{4}\abs{\bra{-}E_i\ket{+}}^2 + \frac{1}{4}\abs{\bra{+}E_i\ket{-}}^2
\end{equation}

From this equation, we find that the estimated Hamming weight of the sampled substring is at least equal to half of the number of applied gates.  For example, applying Pauli $X$ gates to all qubits gives,

\begin{equation}
   M\omega = \frac{M}{N} \sum_i^N \frac{1}{4}\abs{\bra{1}X\ket{0}}^2 + \frac{1}{4}\abs{\bra{0}X\ket{1}}^2 + \frac{1}{4}\abs{\bra{-}X\ket{+}}^2 + \frac{1}{4}\abs{\bra{+}X\ket{-}}^2
\end{equation}

From this we find that the relative Hamming weight of these applied $X$-gates is,

\begin{equation}
    \omega = \frac{1}{2}
\end{equation}

We obtain the same results from applying $Z$-gates as well. $Y$-gates can be detected from every sampling qubit and give a larger relative Hamming weight of $\omega = 1$. Therefore, given the adversary is restricted to Pauli gates, we can estimate the number of gates applied to the message qubits as twice the estimated Hamming weight, $2M\omega$.  Given that the error-correcting code can correct up to $\frac{d-1}{2}$ errors, the code can remove Eve's influence if,

\begin{equation}
    2M\omega \leq \frac{d-1}{2}
\end{equation}

\begin{center}
    \includegraphics[scale = .35]{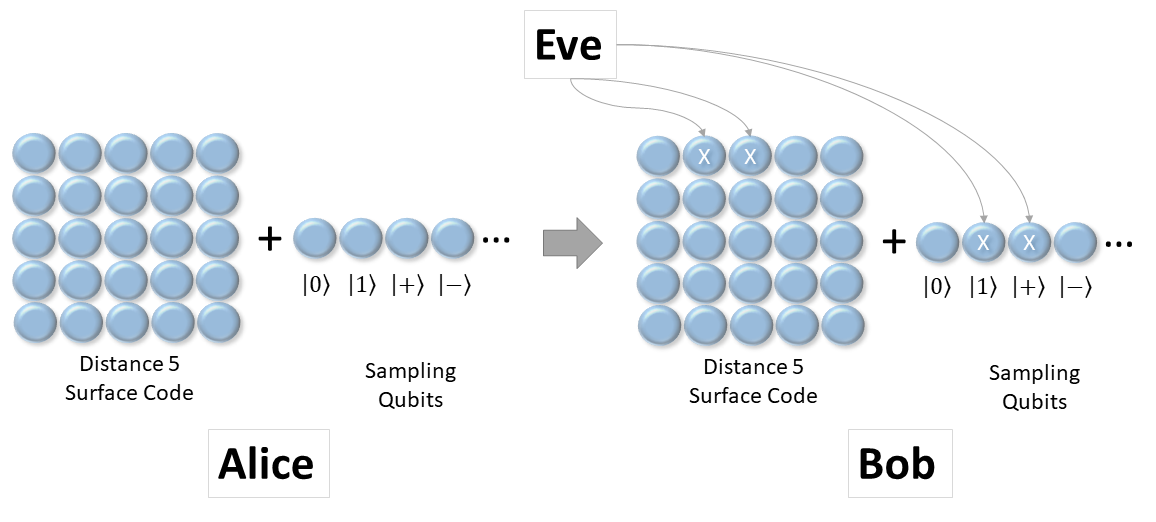}

    \textbf{\footnotesize Figure 2. The proposed protocol using the distance 5 surface code and some sampling qubits.  In this depiction, Eve applies four X-gates in the quantum channel.  Bob uses the sampling qubits to estimate the number of gates applied to the logical qubit and uses the surface code to remove Eve's influence.}
\end{center}

For a specific example, suppose that Alice prepared the logical qubit in the distance 5 surface code as is depicted in Figure 2.  She sends this logical qubit through a quantum channel along with some sampling qubits. After passing through the quantum channel, Bob is able to use the sampling qubits to estimate the Hamming weight of the logical qubit as $M\omega = 1$.  In this scenario, we can use this Hamming weight to estimate that Eve has applied 2 Pauli gates to the logical qubit.  Through this, Bob can determine that the distance 5 surface code prepared by Alice will be able to remove these two gates applied by Eve.

\subsection{Using the Error Bound $\epsilon_{qu}^\delta$}

We must now address the probability that the sampling procedure has failed and Eve has by chance avoided the sampling qubits.  This failure would allow for the true relative Hamming weight to be greater than $\omega + \delta$ with probability $\epsilon_{qu}^\delta$.  This would in turn imply that Eve has applied greater than $2M(\omega + \delta)$ Pauli gates to the logical qubit.  To calculate this failure probability we must first determine the value of $\delta$ which would cause the protocol to fail.  As we are removing Eve's influence through an error-correcting code, the greatest value of $\delta$ we can allow is the value which saturates the error-correcting code.  In this way, $\delta$ is set such that,

\begin{equation}
    2M(\omega+\delta) = \frac{d-1}{2}
\end{equation}

The true relative Hamming weight is less than this estimate of $\omega + \delta$ with probability $1-\epsilon_{qu}^\delta$.  Recall from Section 2 that $\epsilon_{qu}^\delta = 2\exp{-\frac{1}{6}\delta^2k}$, where $k$ is the number of sampling qubits, giving,

\begin{equation}
    1-\epsilon_{qu}^\delta = 1 - 2\exp{-\frac{1}{6M^2}(\frac{d-1}{4}-M\omega)^2k}
\end{equation}

Let us illustrate this with a numerical example.  Alice sends the distance 5 surface code, along with 20000 sampling qubits to Bob.  Bob then performs quantum sampling and estimates the Hamming weight of the message qubits as $M\omega = \frac{1}{2}$.  Since the employed code can correct up to $\frac{5-1}{2} = 2$ errors,  Bob can now perform error correction, keep the logical qubit, and state that the protocol has succeeded with probability,

\begin{equation}
    1 - 2\epsilon_{qu}^\delta = 1 - 2\exp{-\frac{1}{12(25)^2}*20000} = 86.1\%
\end{equation}

As can be seen through this example, this one-way protocol requires significantly more resources to distribute a single entangled qubit than two-way purification protocols.  However, only one round of communication is necessary, decreasing the latency of the protocol at the cost of requiring more qubits.

\section{Security Against Qubit Measurements}
Now that the security of the proposed protocol has been established against single qubit Pauli gates, we can explore the security of the protocol if we additionally allow Eve the ability to perform qubit measurements.  

For example, consider a measurement resulting in finding the qubit in the $\ket{0}$ state,

\begin{equation}
    E_i = M_{\ket{0}} = \ket{0}\bra{0}
\end{equation}

We obtain from Equation 7, 

\begin{equation}
   N\omega = \sum_i^N \frac{1}{4} (\abs{\bra{1}\ket{0}\bra{0}\ket{0}}^2 + \abs{\bra{0}\ket{0}\bra{0}\ket{1}}^2 + \abs{\bra{-}\ket{0}\bra{0}\ket{+}}^2 + \abs{\bra{+}\ket{0}\bra{0}\ket{-}}^2)
\end{equation}

This leads to the Hamming weight,

\begin{equation}
    \omega = \frac{1}{4}
\end{equation}

This similarly follows for measurements $M_{\ket{1}}$ which result in $\ket{1}$.  Qubit measurements therefore increase the Hamming weight by $\frac{1}{4}$.  This makes measurements more difficult to detect with quantum sampling compared to a Pauli gate.

Equation 17 indicates that if we restrict Eve to measurements she can now affect twice as many qubits while retaining the same Hamming weight as compared to Section 4.  For a code with distance $d$, we must ensure that the distance is large enough to correct double the number of tampered qubits.  This changes the requirement in step 6 of the protocol from $2M\omega \leq \frac{d-1}{2}$ to $4M\omega \leq \frac{d-1}{2}$ and Bob's calculation of $\delta$ to,

\begin{equation}
    4M(\omega+\delta) = \frac{d-1}{2}
\end{equation}

\newpage

This changes the probability of success to,

\begin{equation}
    1- 2\epsilon_{qu}^\delta = 1 - 2\exp{-\frac{1}{6M^2}(\frac{d-1}{8}-M\omega)^2k}
\end{equation}

By changing this constraint in the protocol, Alice and Bob will now be able to additionally guarantee security against qubit measurements.

\section{Conclusion and Future Directions}
In this paper, we have proposed a one-way entanglement purification protocol with quantum sampling and proved its security against a restricted adversary with access to Pauli gates and measurements.  The proof of security of this protocol is straightforwardly proven from the properties of quantum error-correcting codes and the sampling framework utilized.

This protocol breaks the symmetry between Bob and Eve which is typically present in one-way entanglement purification protocols by utilizing a shared secret key. This secret key allows Bob to securely conduct quantum sampling without requiring him to send message back to Alice.  This protocol may allow for lower latency as compared to two-way error-correcting protocols, as less communication is needed between the two parties.  However, more qubits are necessary in one-way protocols to achieve similar performance to two-way protocols.

The protocol proposed in this paper has so far only been proven secure to a significantly restricted adversary.  Further research directions include examining the security of this protocol with respect to an adversary with access to arbitrary single or multi-qubit gates.  At first glance, this approach may seem futile when Eve is given access to infinitesimally small gates, as Eve could apply infinitely small rotation gates to each qubit while maintaining approximately zero Hamming weight.  However, two facts help to mitigate the effectiveness of this approach for Eve.  First, error-correcting codes discretize errors, and therefore would discretize the small gates Eve has applied, suppressing them from affecting the logical qubit.  Second, the Eastin Knill theorem states that an operator made of infinitesimally small transverse gates cannot be a fault-tolerant logical operator.  Due to this, any single qubit gates Eve employs must be finite sized to manipulate the logical qubit sent by Alice \cite{Eastin_2009}.  This indicates that there may be a lower limit to the size of single qubit gates which Eve can apply to improve her likelihood of eavesdropping.  

However, as the Eastin-Knill theorem does not apply to multi-qubit non-transverse gates, it may be more difficult to prove the security of one-way entanglement purification against arbitrary multi-qubit attacks.  This would require research into the lowest Hamming weight multi-qubit operator that can perform logical rotations on an error corrected qubit.  However, there has been relatively little research into constructing continuous logical operators on error-corrected qubits \cite{Eastin_2009, cianci2023constructing}, since fault-tolerant quantum computation can be achieved with finite sized gate sets such as Clifford+T \cite{nielsen00}.  Therefore, further examination into constructing continuous logical operators may have applications in the security of this protocol against arbitrary multi-qubit attacks.

\printbibliography

\end{document}